\documentclass{kapproc}  

\usepackage{t1enc}       
\usepackage{procps} 
\usepackage[dvips]{graphicx}
\usepackage{bm}          
\usepackage{amssymb}     

\chapsectrunningheads    
\upperandlowercase       
\normallatexbib          

\begin{document}
\def\bea{\begin{eqnarray}}
\def\eea{\end{eqnarray}}
\def\be{\begin{equation}}
\def\ee{\end{equation}}
\def\rra{\right\rangle}
\def\lla{\left\langle}
\def\sig{\sigma}
\def\eps{\epsilon}
\def\sgm{\Sigma^-}
\def\la{\Lambda}
\def\pv{\bm{p}}
\def\tv{\bm{\tau}}
\def\sv{\bm{\sigma}}
\def\rv{\bm{r}}
\def\sdot{\!\cdot\!}
\def\tt{(\tv_i\sdot\tv_j)}
\def\ss{(\sv_i\sdot\sv_j)}
\def\ns{neutron star}
\def\dir{/home/axparct/users1/schulze/me/tex/pub}

\articletitle{Neutron Star Structure with Hyperons and Quarks}

\author{M. Baldo, F. Burgio, H.-J. Schulze}
\affil{INFN Sezione di Catania, Via S. Sofia 64, I-95123 Catania, Italy}

\begin{abstract}
We discuss the high-density nuclear equation of state 
within the Brueckner-Hartree-Fock approach.
Particular attention is paid to the effects of 
nucleonic three-body forces, the presence of hyperons,
and the joining with an eventual quark matter phase.
The resulting properties of neutron stars,
in particular the mass-radius relation,
are determined.
It turns out that stars heavier than 1.3 solar masses contain
necessarily quark matter.
\end{abstract}


\section{Brueckner theory}

Over the last two decades the increasing interest for the equation
of state (EOS) of nuclear matter has stimulated a great deal of
theoretical activity.  
Phenomenological and microscopic models of
the EOS have been developed along parallel lines with
complementary roles.
The former models include nonrelativistic
mean field theory based on Skyrme interactions \cite{bon} and
relativistic mean field theory based on meson-exchange
interactions (Walecka model) \cite{wal}. 
Both of them fit the parameters of the interaction in order 
to reproduce the empirical saturation properties of nuclear matter 
extracted from the nuclear mass table. 
The latter ones include nonrelativistic
Brueckner-Hartree-Fock (BHF) theory \cite{bal} and its relativistic
counterpart, the Dirac-Brueckner (DB) theory \cite{dbhf}, the
nonrelativistic variational approach also corrected by
relativistic effects \cite{pan}, and more recently the chiral
perturbation theory \cite{chi}. 
In these approaches the parameters
of the interaction are fixed by the experimental nucleon-nucleon
and/or nucleon-meson scattering data. 

For states of nuclear matter
with high density and high isospin asymmetry the experimental
constraints on the EOS are rather scarse and indirect. 
Different approaches lead to different or even contradictory
theoretical predictions for the nuclear matter properties.
The interest for these properties lies, to a large extent, 
in the study of astrophysical objects, i.e.,
supernovae and neutron stars. 
In particular, the structure of a neutron star is very sensitive 
to the compressibility and the symmetry energy. 
The neutron star mass, measured in binary systems, 
has been proposed as a constraint for the EOS of nuclear
matter \cite{gle}.

One of the most advanced microscopic approaches to the EOS of 
nuclear matter is the Brueckner theory. 
In the recent years, it has made a rapid
progress in several aspects:
(i) The convergence of the
Brueckner-Bethe-Goldstone (BBG) expansion has been firmly
established \cite{son,thl}. 
(ii) Important relativistic effects have been
incorporated by including into the interaction the virtual
nucleon-antinucleon excitations, and the relationship with the DB
approach has been numerically clarified \cite{gra}. 
(iii) The addition of microscopic three-body forces (TBF) based on nucleon
excitations via pion and heavy meson exchanges, permitted to
improve to a large extent the agreement with the empirical
saturation properties \cite{gra,mic,lom}.
(iv) Finally, the BHF approach has been extended in a fully microscopic and 
self-consistent way to describe nuclear matter containing also 
hyperons \cite{hypmat}, 
opening new fields of applications such as hypernuclei \cite{hypnuc}
and a more realistic modeling of neutron stars \cite{hypns,barc}.

In the present article we review these issues and
present our results for neutron star structure 
based on the resulting EOS of dense hadronic matter.

\subsubsection{Convergence of the hole-line expansion}

\begin{figure} 
\includegraphics[scale=0.58,bb=70 60 400 60]{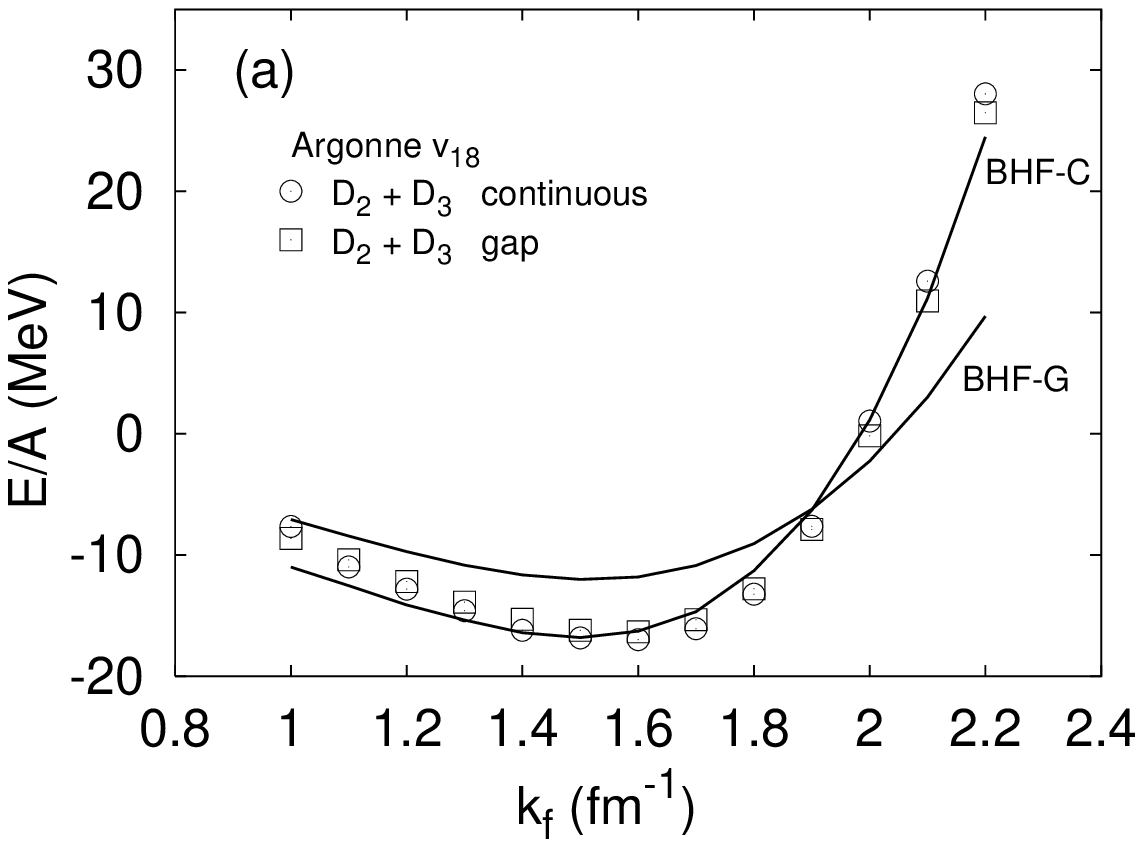}
\includegraphics[scale=0.39,angle=90,bb=185 580 535 580]{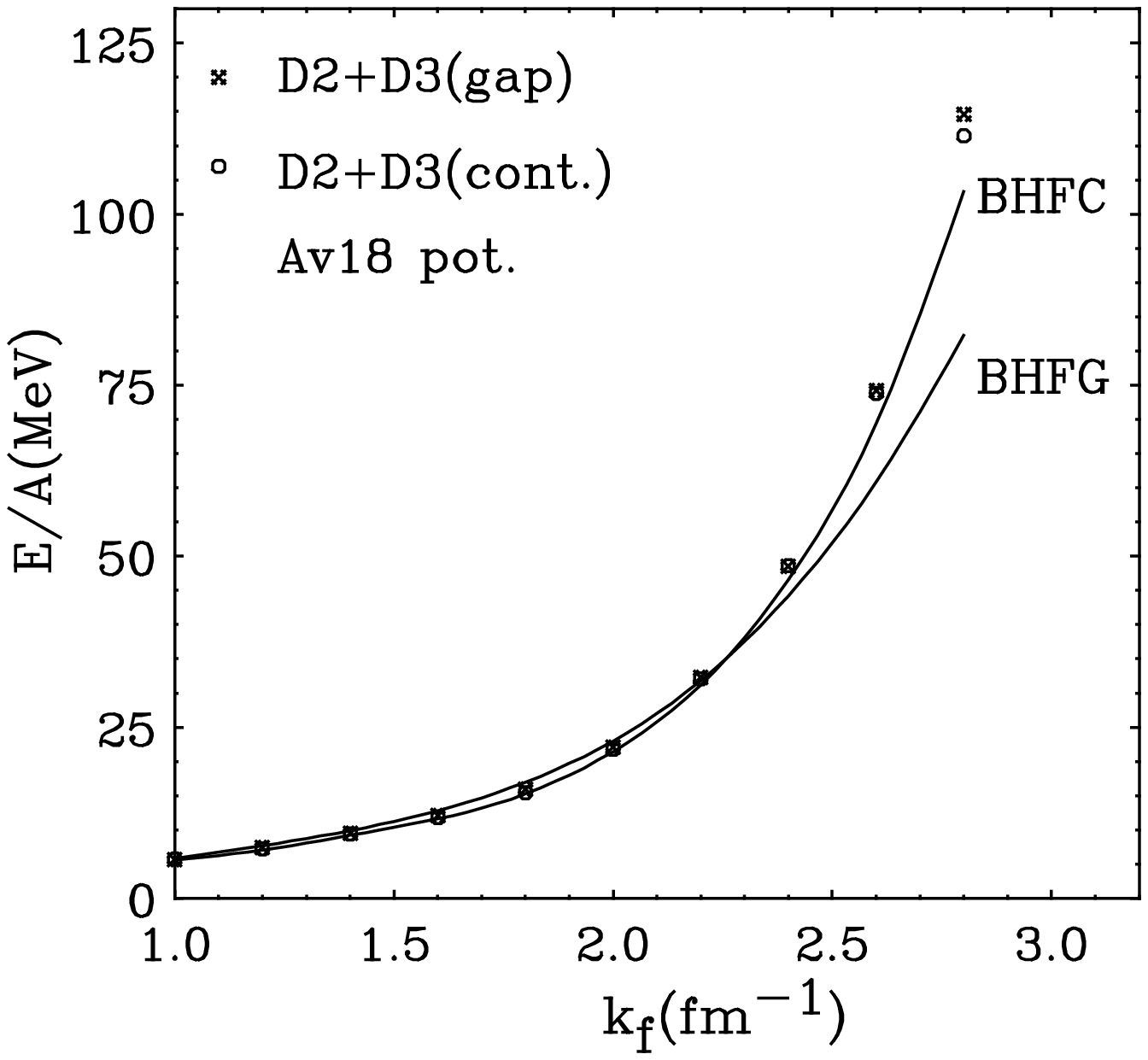}
\caption{
Comparison of BHF two hole-line (lines) and three hole-line (markers) 
results for symmetric nuclear matter (left plot) 
and pure neutron matter (right plot),
using continuous and gap choice for the single-particle potentials.}
\label{f:ba}
\end{figure}

The nonrelativistic BBG expansion of the nuclear matter correlation energy
$E/A$ can be cast as a power series in terms of the number
of hole lines contained in the corresponding diagrams, which
amounts to a density power expansion \cite{bal}. 
The two hole-line truncation is named the Brueckner-Hartree-Fock (BHF)
approximation. 
At this order the energy $D_2$ is very much
affected by the choice of the auxiliary potential, as shown in
Fig.~\ref{f:ba} (solid lines), 
where the numerical results obtained
with the gap and the continuous choice are compared
for symmetric nuclear matter as well as neutron matter.
But, as also shown in the same figure \cite{son,thl},
adding the three-hole line contributions $D_3$, the resulting EOS
is almost insensitive to the choice of the auxiliary potential,
and very close to the result $D_2$ with the continous choice.

In spite of the satisfactory convergence, the saturation density misses the
empirical value $\rho_0=0.17\;{\rm fm}^{-3}$
extracted from the nuclear mass tables. 
This confirms the belief that the concept of a many nucleon system
interacting with only a two-body force is not adequate to
describe nuclear matter, especially at high density.

\subsubsection{Relativistic corrections}

\begin{figure} 
\includegraphics[scale=0.85,bb=100 310 400 540]{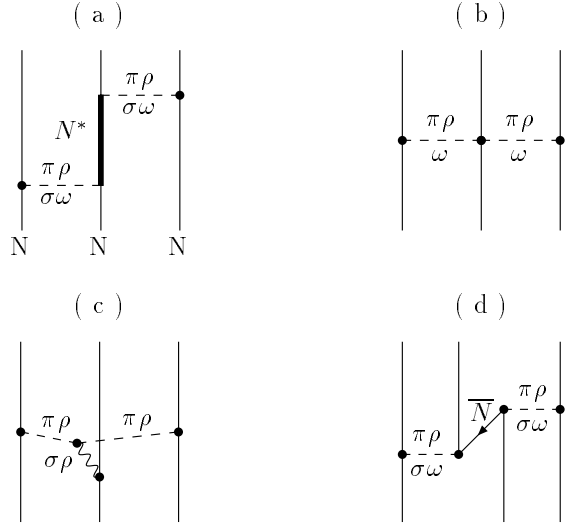}
\caption{
Various diagrams contributing to the microscopic TBF.}
\label{f:dia}
\end{figure}

Before the possible effects of TBF are examined, 
one should introduce relativistic corrections in the
preceding nonrelativistic BHF predictions. 
This is done in the Dirac-Brueckner approach \cite{dbhf},
where the nucleons, instead of propagating as plane waves,
propagate as spinors in a mean field with a scalar component $U_S$
and a vector component $U_V$, 
self-consistently determined together with the $G$-matrix. 
The nucleon self-energy
can be expanded in terms of the scalar field,
\be
 \Sigma(\pv) = \sqrt{(M+U_S)^2+\pv^2} + U_V 
 \approx 
 p_0 + U_V + \frac{M}{p_0}U_S + \frac{\pv^2}{2p_0^3} U_S^2 + \ldots \:.
\ee
The second-order term can be interpreted as due to the
interaction between two nucleons with the virtual excitation of a
nucleon-antinucleon pair \cite{brow}. 
This interaction is a TBF with the exchange of a scalar ($\sigma$) meson, 
as illustrated by the diagram (d) of Fig.~\ref{f:dia}.
Actually this diagram represents a class of
TBF with the exchange of light ($\pi$, $\rho$) and heavy ($\sigma,\omega$) mesons. 
There are, however, several other diagrams representing TBF,
Fig.~\ref{f:dia}(a-c),
which should be evaluated as well in a consistent treatment of TBF.

\section{Three-body forces}

Since long it is well known that two-body forces are not enough to
explain some nuclear properties, and TBF have to be introduced.
Typical examples are: the binding energy of light nuclei, the spin
dynamics of nucleon-deuteron scattering, and the saturation point
of nuclear matter. 
Phenomenological and microscopic TBF have been
widely used to describe the above mentioned properties. 

In the framework of the Brueckner theory a rigorous treatment of TBF
would require the solution of the Bethe-Faddeev equation,
describing the dynamics of three bodies embedded in the nuclear matter. 
In practice a much simpler approach is employed, namely
the TBF is reduced to an effective,
density dependent, two-body force by averaging over the third
nucleon in the medium, taking account of the nucleon-nucleon
correlations by means of the BHF defect function $g_{ij}$,
\be
 \lla 12|V(\rho)|1'2' \rra = \sum_{33'}
 \Psi^*_{123} \lla 123|V|1'2'3' \rra \Psi_{1'2'3'} \:.
\ee
Here $\Psi_{123}=\phi_3(1-g_{13})(1-g_{23})$ and $\phi_3$ is the
free wave function of the third particle. 
This effective two-body
force is added to the bare two-body force and recalculated at each
step of the iterative procedure.

\subsubsection{Microscopic TBF}

The microscopic TBF of Refs.~\cite{gra,mic,lom} is based on 
meson-exchange mechanisms accompanied by the excitation of
nucleonic resonances,
as represented by the diagrams plotted in Fig.~\ref{f:dia}.
Besides the TBF arising from the excitation of a $N\bar N$ pair [diagram (d)],
already discussed in the preceding section,
another important class of TBF [diagram (a)]
is due to the excitation of the isobar $\Delta(1232)$ resonance 
via the exchange of light ($\pi$, $\rho$) mesons,
or the lowest non-isobar nucleon excitation $N^*(1440)$
excited by heavy meson ($\sigma$ and $\omega$) exchanges.
Diagrams (b) and (c) are included only for completeness
and play a minor role \cite{gra}. 

The combined effect of these TBF
is a remarkable improvement of the saturation properties
of nuclear matter \cite{lom}. 
Compared to the BHF prediction with only two-body forces, 
the saturation energy is shifted from $-18$ to $-15$ MeV,
the saturation density from 0.26 to 0.19 $\rm fm^{-3}$, 
and the compression modulus from 230 to 210 MeV. 
The spin and isospin properties with TBF 
exhibit also quite satisfactory behaviour \cite{land}.

\subsubsection{Phenomenological TBF}

A second class of TBF that are widely used in the literature, in particular
for variational calculations of finite nuclei and nuclear matter \cite{pan},
are the phenomenological Urbana TBF \cite{uix}.
We remind that the Urbana IX TBF model contains a two-pion
exchange potential $V_{ijk}^{2\pi}$ supplemented by a phenomenological
repulsive term $V_{ijk}^R$,
\bea
  V_{ijk} = V_{ijk}^{2\pi} + V_{ijk}^R \:,
\eea
where
\bea
 V_{ijk}^{2\pi}  \!\!\!&=&\!\!\! 
 A \sum_{\rm cyc} \bigg[\!
 \left\{X_{ij},X_{jk}\right\} 
 \left\{ \tv_i\sdot\tv_j, \tv_j\sdot\tv_k \right\}
 + {1\over 4} \left[X_{ij},X_{jk}\right] 
 \left[ \tv_i\sdot\tv_j, \tv_j\sdot\tv_k \right]
 \!\bigg] \:,\phantom{aaa}
\\
 V_{ijk}^R  \!\!\!&=& \!\!\! 
 U \sum_{\rm cyc} T^2(m_\pi r_{ij}) T^2(m_\pi r_{jk}) \:.
\eea
The two-pion exchange operator $X_{ij}$ is given by
\be
  X_{ij} = Y(m_\pi r_{ij}) \sv_i\sdot\sv_j
  + T(m_\pi r_{ij}) S_{ij} \:,
\ee 
where 
$S_{ij}=3(\sv_i \sdot \hat{\rv}_{ij})(\sv_j \sdot \hat{\rv}_{ij}) 
- \sv_i\sdot\sv_j$ 
is the tensor operator and
$\sv$ and $\tv$ are the Pauli spin and isospin operators.
$Y$ and $T$ are the Yukawa and tensor functions, respectively, 
associated to the one-pion exchange \cite{uix}. 

After reducing this TBF to an effective, density dependent, 
two-body force by the averaging procedure described earlier, 
the resulting effective two-nucleon potential assumes a simple structure, 
\be
 \overline{V}_{\!\!ij}^{\rm pheno}(\rv) = 
 \tt \Big[ \ss V^{2\pi}_C(r) + S_{ij}(\hat{\rv}) V^{2\pi}_T(r) \Big]
  + V^{R}(r) \:,
\label{e:vavp}
\ee
containing central and tensor two-pion exchange components as well as
a central repulsive contribution.
For comparison, the averaged microscopic TBF \cite{gra}
involves five different components:
\bea
  \overline{V}_{\!\!ij}^{\rm micro}(\rv) &=&
  \tt\ss V^{\tau\sig}_C(r) + \ss V^{\sig}_C(r) + V_C(r)
\nonumber\\ &&
  +\, S_{ij}(\hat{\rv}) \Big[ \tt V^{\tau}_T(r) + V_T(r) \Big] \:.
\label{e:vavm}
\eea

In the variational approach the two parameters $A$ and $U$ are determined by
fitting the triton binding energy together with the saturation density
of nuclear matter
(yielding however too little attraction,
$E/A\approx-12\;\rm MeV$,
in the latter case \cite{pan}).
In the BHF calculations they are instead chosen to reproduce the empirical 
saturation density together with the binding energy of nuclear matter.
The resulting parameter values are
$A = -0.0293\;{\rm MeV}$ and  $U = 0.0048\;{\rm MeV}$
in the variational Urbana IX model, whereas for the optimal BHF+TBF
calculations 
we require $A = -0.0333\;{\rm MeV}$ and  $U = 0.00038\;{\rm MeV}$,
yielding a saturation point at $k_F\approx 1.36\;{\rm fm}^{-1}$,
$E/A \approx -15.5\;{\rm MeV}$, and an incompressibility
$K \approx 210\;{\rm MeV}$.

These values of $A$ and $U$ have been obtained by using the Argonne $v_{18}$
two-body force \cite{v18} 
both in the BHF and in the variational many-body theories.
However, the required repulsive component ($\sim U$)
is much weaker in the BHF approach,
consistent with the observation that in the variational calculations
usually heavier nuclei as well as nuclear matter are underbound.
Indeed, less repulsive TBF became available recently \cite{ill} in order
to address this problem.

\begin{figure} 
\includegraphics[scale=0.45,angle=270,bb=170 35 535 35]{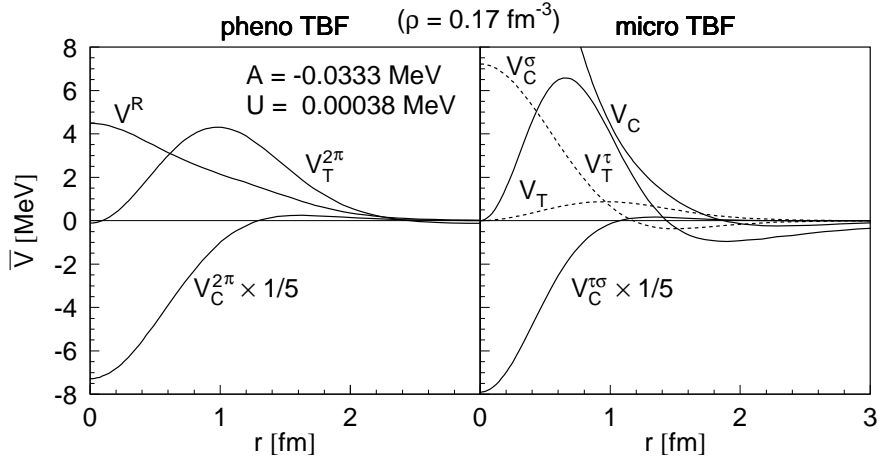}
\caption{
Comparison of the different components of averaged
phenomenological and microscopic TBF, 
Eqs.~(\ref{e:vavp}) and (\ref{e:vavm}).}
\label{f:vtbf}
\end{figure} 

\section{EOS of nuclear matter from 
different TBF}

In Fig.~\ref{f:vtbf} we compare the different
components 
$V^{2\pi}_C, V^{2\pi}_T, V^R$, Eq.~(\ref{e:vavp}),
and 
$V^{\tau\sig}_C, V^\sig_C, V_C, V^\tau_T, V_T$, Eq.~(\ref{e:vavm}), 
of the averaged phenomenological and microscopic TBF potentials in 
symmetric matter at normal density. 
One notes that the attractive components 
$V^{2\pi}_C, V^{2\pi}_T$ 
and 
$V^{\tau\sig}_C, V^\tau_T$ 
roughly correspond to each other, 
whereas the repulsive part ($V^R$ vs.~$V_C$) 
is much larger for the microscopic TBF. 
With the choice of parameters $A$ and $U$
given above, one would therefore expect a more repulsive
behaviour of the microscopic TBF, which is indeed confirmed in the
following.

\begin{figure} 
\includegraphics[scale=0.57,angle=270,bb=155 47 474 635]{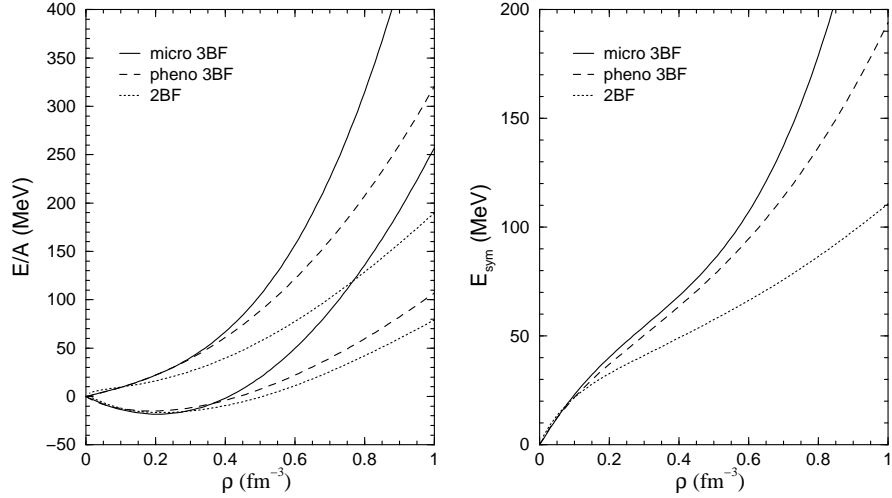}
\caption{
Left plot:
Binding energy per nucleon of symmetric nuclear matter
(lower curves using a given linestyle) and pure neutron matter
(upper curves), employing different TBF.
Right plot:
Corresponding symmetry energy of nuclear matter.}
\label{f:sym}
\end{figure}

Let us now confront the EOS predicted by the phenomenological
TBF and the microscopic one. 
In both cases the BHF
approximation has been adopted with same two-body force 
(Argonne $v_{18}$).
In the left panel of Fig.~\ref{f:sym} we display the equation of state 
both for symmetric matter
(lower curves) and pure neutron matter (upper curves).
We show results obtained for several cases, i.e.,
i) only two-body forces are included (dotted lines),
ii) TBF implemented within the phenomenological Urbana IX model 
(dashed lines), and
iii) TBF treated within the microscopic meson-exchange approach 
(solid lines).
We notice that the EOS for symmetric matter with TBF
reproduces the correct nuclear matter saturation point.
Moreover, the incompressibility turns out to be compatible
with the values extracted from phenomenology, i.e.,
$K \approx 210\;\rm MeV$.
Up to a density of $\rho \approx 0.4\;{\rm fm}^{-3}$
the microscopic and phenomenological TBF are in fair agreement,
whereas at higher density the microscopic TBF turn out to be more repulsive.

Within the BHF approach, 
it has been verified \cite{hypns,bom} 
that a parabolic approximation for the binding
energy of nuclear matter with arbitrary proton fraction $x$ 
is well fulfilled,
\be
 {E\over A}(\rho,x) \approx 
 {E\over A}(\rho,x=0.5) + (1-2x)^2 E_{\rm sym}(\rho) \:,
\label{e:parab}
\ee
where
the symmetry energy $E_{\rm sym}$ can be expressed in
terms of the difference of the energy per particle between pure neutron 
($x=0$) and symmetric ($x=0.5$) matter:
\be
  E_{\rm sym}(\rho) = 
  - {1\over 4} {\partial(E/A) \over \partial x}(\rho,0)
  \approx {E\over A}(\rho,0) - {E\over A}(\rho,0.5) \:.
\label{e:sym}
\ee
In the right panel of Fig.~\ref{f:sym} we display the symmetry energy as a
function of the nucleon density $\rho$ for different choices of the TBF.
We observe results in agreement with the characteristics of the
EOS shown in the left panel.
Namely, the stiffest equation of state, i.e.,
the one calculated with the microscopic TBF,
yields larger symmetry energies compared to the ones obtained with the
Urbana phenomenological TBF.
Moreover, the symmetry energy calculated 
(with or without TBF) 
at the saturation point yields a value $E_{\rm sym}\approx 30\;\rm MeV$,
compatible with nuclear phenomenology.

\section{Neutron star structure}

In order to study the effects of different TBF on neutron star
structure, we have to calculate the composition and the EOS
of cold, catalyzed matter. 
We require that the neutron star contains charge neutral matter 
consisting of neutrons, protons, and leptons ($e^-$, $\mu^-$)
in beta equilibrium. 
Using the various TBF discussed above,
we compute the proton fraction and the EOS for 
charge neutral and beta-stable matter
in the following standard way \cite{ns,aa}:
The Brueckner calculation yields the energy density of 
lepton/baryon matter 
as a function of the different partial densities, 
\bea
 \eps(\rho_n,\rho_p,\rho_e,\rho_\mu) &=&
 (\rho_n m_n +\rho_p m_p) + (\rho_n+\rho_p) {E\over A}(\rho_n,\rho_p)
\nonumber\\
 && + \rho_\mu m_\mu + {1\over 2m_\mu}{(3\pi^2\rho_\mu)^{5/3} \over 5\pi^2}
\nonumber\\
 && + { (3\pi^2\rho_e)^{4/3} \over 4\pi^2} \:,
\label{e:epsnn}
\eea
where we have used ultrarelativistic and nonrelativistic approximations
for the energy densities of electrons and muons, respectively.
The various chemical potentials 
(of the species $i=n,p,e,\mu$)
can then be computed straightforwardly,
\be
 \mu_i = {\partial \eps \over \partial \rho_i} \:,
\ee
and the equations for beta-equilibrium,
\be
\mu_i = b_i \mu_n - q_i \mu_e \:,
\ee 
($b_i$ and $q_i$ denoting baryon number and charge of species $i$)
and charge neutrality,
\be 
 \sum_i \rho_i q_i = 0 \:,
\ee
allow to determine the
equilibrium composition $\rho_i(\rho)$
at given baryon density $\rho=\rho_n+\rho_p$
and finally the EOS,
\be
 p(\rho) = \rho^2 {d\over d\rho} 
 {\eps(\rho_i(\rho))\over \rho}
 = \rho {d\eps \over d\rho} - \eps 
 = \rho \mu_n - \eps \:.
\ee

In order to calculate the mass-radius relation, one has then to solve the
well-known Tolman-Oppenheimer-Volkov equations \cite{ns}, 
\bea
 {dp \over dr} &=& - { G m \over r^2}\, 
 { (\eps+p)(1+4\pi r^3 p/m) \over  1 - 2Gm/r} \:,
\\
 { dm \over dr} &=& 4\pi r^2 \eps \:,
\eea
with the newly constructed EOS for the charge neutral and beta-stable case
as input
(supplemented by the EOS of
Feynman-Metropolis-Teller \cite{fey}, Baym-Pethick-Sutherland \cite{baym}, 
and Negele-Vautherin \cite{nv}
for the outer part of the neutron star,
$\rho \lesssim 0.08\;{\rm fm}^{-3}$).
The solutions provide information on the interior structure of a star, 
$\rho(r)$, as well as the mass-radius relation, $M(R)$.

\begin{figure} 
\includegraphics[scale=0.58,angle=270,bb=153 56 480 56]{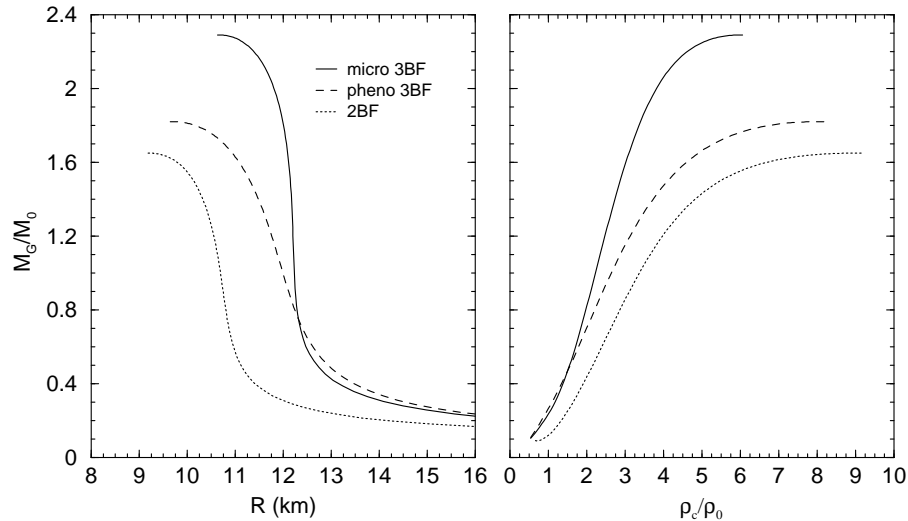}
\caption{
The neutron star gravitational mass (in units of solar mass $M_\odot$)
is displayed vs.~the radius (left panel) 
and the normalized central baryon density $\rho_c$ 
($\rho_0=0.17\;{\rm fm}^{-3}$) (right panel).}
\label{f:mr}
\end{figure} 

The results are shown in Fig.~\ref{f:mr}.
We notice that the EOS calculated with the microscopic TBF produces
the largest gravitational masses, with the maximum mass of the order of
2.3 $\rm M_\odot$,
whereas the phenomenological TBF yields a maximum mass of
about 1.8 $\rm M_\odot$.
In the latter case, neutron stars are characterized by smaller radii and
larger central densities, i.e.,
the Urbana TBF produce more compact stellar objects.
For completeness, we also show a sequence of stellar configurations obtained
using only two-body forces.
In this case the maximum mass is slightly above 1.6 $M_\odot$,
with a radius of 9 km and a central density equal to 9 times
the saturation value.

However, these results should be considered as only provisory,
since it is well known that the inclusion of hyperons \cite{hypns,barc}
or quark matter \cite{qm} may strongly affect the structure of the star, 
in particular reducing substantially the maximum mass.
We discuss this point now in detail.

\section{Hyperons in nuclear matter}

While at moderate densities $\rho \approx \rho_0$ the matter inside 
a neutron star consists only of nucleons and leptons, 
at higher densities several other
species of particles may appear due to the fast rise of the baryon 
chemical potentials with density. 
Among these new particles are strange baryons, namely, 
the $\Lambda$, $\Sigma$, and $\Xi$ hyperons. 
Due to its negative charge, the $\Sigma^-$ hyperon is the 
first strange baryon expected to appear with increasing density in the 
reaction $n+n \rightarrow p+\Sigma^-$,
in spite of its substantially larger mass compared to the neutral 
$\Lambda$ hyperon ($M_{\Sigma^-}=1197\;{\rm MeV}, M_\Lambda=1116\;{\rm MeV}$).
Other species might appear in stellar matter,
like $\Delta$ isobars along with pion and kaon condensates.
It is therefore mandatory to generalize the study of the nuclear EOS
with the inclusion of the possible hadrons, other than nucleons, which
can spontaneously appear in the inner part of a neutron star, just because their
appearance is able to lower the ground state energy of the nuclear matter
dense phase. 
In the following we will concentrate on the production 
of strange baryons and assume that a baryonic description of nuclear matter 
holds up to densities as those encountered in the core of neutron stars. 

As we have seen in the previous sections, the nuclear EOS can be 
calculated with good accuracy in the Brueckner two hole-line 
approximation with the continuous choice for the single-particle
potential, since the results in this scheme are quite close to the full
convergent calculations which include also the three hole-line
contribution. 
It is then natural to include the hyperon degrees of freedom
within the same approximation to calculate the nuclear EOS needed
to describe the neutron star interior. 
To this purpose, one requires in principle 
nucleon-hyperon (NY) and hyperon-hyperon (YY) potentials. 
In our work we use the Nijmegen soft-core NY potential \cite{mae89}
that is well adapted to the available experimental NY scattering data.
Unfortunately, up to date no YY scattering data 
and therefore no reliable YY potentials are available.
We therefore neglect these interactions in our calculations,
which is supposedly justified, as long as the hyperonic partial 
densities remain limited.
Also, for the following calculations the $v_{18}$ NN potential together
with the phenomenological TBF introduced previously, are used.

With the NN and NY potentials, the various $G$ matrices are evaluated 
by solving numerically the Brueckner equation, which can be written in 
operatorial form as \cite{hypmat,hypns}
\be
  G_{ab}[W] = V_{ab} + \sum_c \sum_{p,p'} 
  V_{ac} \Big|pp'\Big\rangle 
  { Q_c \over W - E_c +i\eps} 
  \Big\langle pp'\Big| G_{cb}[W] \:, 
\label{e:g}
\ee
where the indices $a,b,c$ indicate pairs of baryons
and the Pauli operator $Q$ and energy $E$ 
determine the propagation of intermediate baryon pairs.
In a given nucleon-hyperon channels $c=(NY)$ one has, for example,
\be
  E_{(NY)} = m_N + m_Y + {k_N^2\over2m_N} + {k_Y^2\over 2m_Y} +
  U_N(k_N) + U_Y(k_Y) \:.
\label{e:e}
\ee
The hyperon single-particle potentials within the continuous choice
are given by
\be
  U_Y(k) = {\rm Re}\, \sum_{N=n,p}\sum_{k'<k_F^{(N)}} 
  \Big\langle k k' \Big| G_{(NY)(NY)}\big[E_{(NY)}(k,k')\big] 
  \Big| k k' \Big\rangle 
\label{e:uy}
\ee
and similar expressions of the form
\be
 U_N(k) = \sum_{N'=n,p} U_N^{(N')}(k) + 
 \sum_{Y=\Sigma^-,\Lambda} U_N^{(Y)}(k) 
\label{e:un}
\ee
apply to the nucleon single-particle potentials.
The nucleons feel therefore direct effects of the other nucleons as well as 
of the hyperons in the environment, whereas for the hyperons there are only 
nucleonic contributions, because of the missing hyperon-hyperon potentials.
The equations (\ref{e:g}--\ref{e:un}) define the BHF scheme with the 
continuous choice of the single-particle energies.  
Due to the occurrence of $U_N$ and $U_Y$ in Eq.~(\ref{e:e}) they constitute 
a coupled system that has to be solved in a self-consistent manner.
In contrast to the standard purely nucleonic calculation
there is now an additional coupled channel structure,
which renders a self-consistent calculation quite time-consuming.
 
\begin{figure}[t] 
\includegraphics[scale=0.445,angle=270,bb=130 27 550 27]{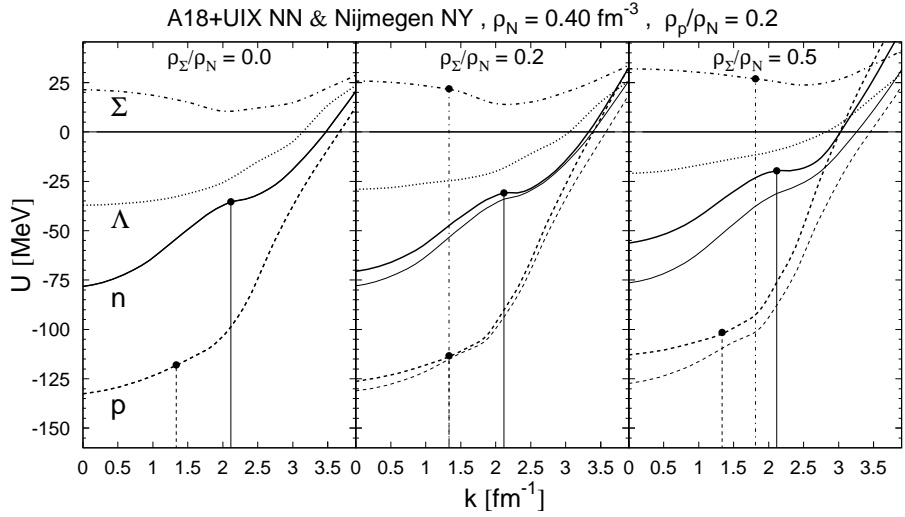}
\caption{
  The single-particle potentials of nucleons $n$, $p$ 
  and hyperons $\sgm$, $\la$ in baryonic matter of fixed nucleonic
  density $\rho_N=0.4\,\rm fm^{-3}$, proton density $\rho_p/\rho_N=0.2$,
  and varying $\sgm$ density $\rho_\Sigma/\rho_N=0.0,0.2,0.5$.
  The vertical lines represent the corresponding Fermi momenta of 
  $n$, $p$, and $\sgm$. 
  For the nucleonic curves, the thick lines represent the complete
  single-particle potentials $U_N$, whereas the thin lines show the values 
  excluding the $\sgm$ contribution, i.e., $U_N^{(n)} + U_N^{(p)}$.}
\label{f:u}
\end{figure} 

Once the different single-particle potentials are known,
the total nonrelativistic baryonic energy density, $\eps$,  
can be evaluated:
\begin{eqnarray}
 \eps &\!=\!& \sum_{i=n,p,\Sigma^-,\Lambda} \int_0^{k_F^{(i)}}\!\! 
 {dk\,k^2\over\pi^2} 
 \left[ m_i + {k^2\over{2m_i}} + {1\over2}U_i(k) \right] 
\\
 \!&=&\! \eps_{NN} +
 \!\!\!\sum_{Y=\sgm,\la} 
 \int_0^{k_F^{(Y)}}\!\!\! {dk\,k^2\over\pi^2}\!
 \left[ m_Y + {k^2\over 2m_Y} + 
 U_Y^{(n)}(k) + U_Y^{(p)}(k) \right] \:, \phantom{aaa}
\label{e:epsny}
\eea
where $\eps_{NN}$ is the nucleonic part of the energy density, Eq.~(\ref{e:epsnn}).
Using for example an effective mass approximation for the hyperon single-particle 
potentials, one could write the last term due to the nucleon-hyperon interaction as
\be
 \eps_{NY} = \sum_{Y=\sgm,\la} \left(
 \rho_Y \Big[ m_Y + U_Y(0) \Big] +
 {1\over 2m_Y^*}{(3\pi^2\rho_Y)^{5/3} \over 5\pi^2} \right) \:,
\ee
which should be added to Eq.~(\ref{e:epsnn}).

The different single-particle potentials involved in the previous equations
are illustrated in Fig.~\ref{f:u},
where neutron and proton densities are fixed,
given by $\rho_N=0.4\,\rm fm^{-3}$ and $\rho_p/\rho_N=0.2$,
and the  $\Sigma^{-}$ density is varied.
Under these conditions the  
$\Sigma^{-}$ single-particle potential is sizeably repulsive, while
$U_\Lambda$ is still attractive (see also Ref.~\cite{hypns}) and the nucleons
are both strongly bound.
The $\sgm$ single-particle potential has a particular shape with an 
effective mass $m^*/m$ close to 1,
whereas the $\la$ effective mass is typically about 0.8 and the
nucleon effective masses are much smaller.

\begin{figure}[t] 
\includegraphics[scale=0.75,angle=270,bb=100 120 480 120]{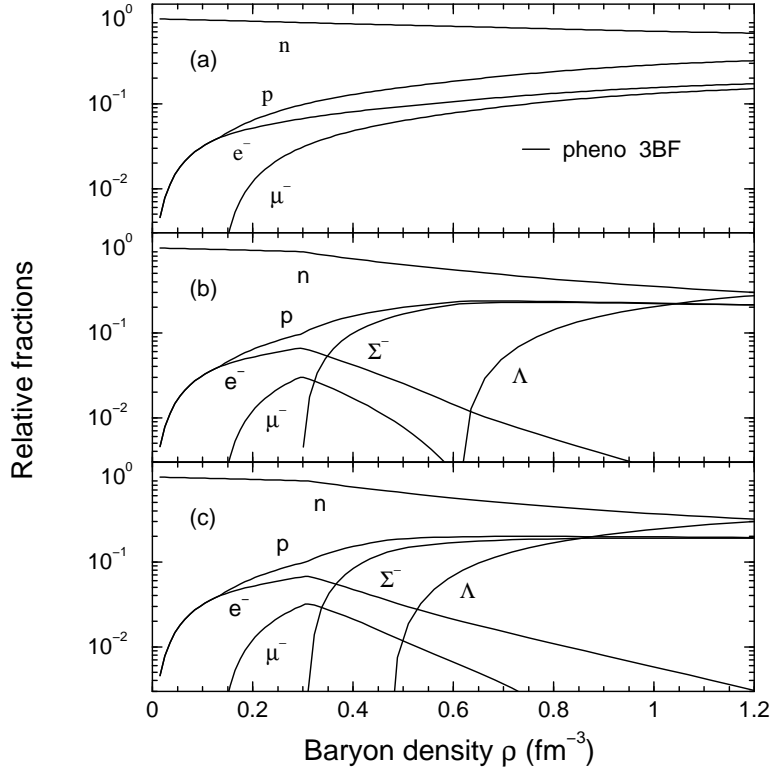}
\caption[]{
The equilibrium composition of asymmetric and $\beta$-stable nuclear matter
comprising
(a) only nucleons,
(b) noninteracting hyperons,
(c) interacting hyperons.
\label{cortona1}}
\end{figure} 

The knowledge of the energy density  
allows then to compute EOS and neutron star structure as described before, 
now making allowance for the species $i=n,p,\sgm,\la,e^-,\mu^-$.
The main physical features of the nuclear EOS which determine the
resulting compositions are essentially the symmetry energy of the nucleon 
part of the EOS and the hyperon single-particle potentials inside 
nuclear matter. 
Since at low enough density the nucleon matter is quite 
asymmetric, the small percentage of protons feel a deep single-particle
potential, and therefore it is energetically convenient to create 
a $\Sigma^{-}$ hyperon, since then a neutron can be converted into a proton.
The depth of the proton potential is mainly determined by the
nuclear matter symmetry energy. 
Furthermore, the potentials felt by the
hyperons can shift substantially the threshold density at which each
hyperon sets in. 

In Fig.~\ref{cortona1} we show the chemical composition of the resulting
$\beta$-stable and asymmetric nuclear matter containing hyperons.
We observe rather low hyperon onset densities of about 2-3 times 
normal nuclear matter density
for the appearance of the $\sgm$ and $\la$ hyperons.
(Other hyperons do not appear in the matter).
Moreover, an almost equal percentage of nucleons and hyperons are 
present in the stellar core at high densities. 
A strong deleptonization of matter takes place, 
since it is energetically convenient 
to maintain charge neutrality through hyperon formation rather than $\beta$-decay.
This can have far reaching consequences for the onset of kaon condensation.

\begin{figure}[t] 
\includegraphics[scale=0.6,angle=270,bb=170 0 450 0]{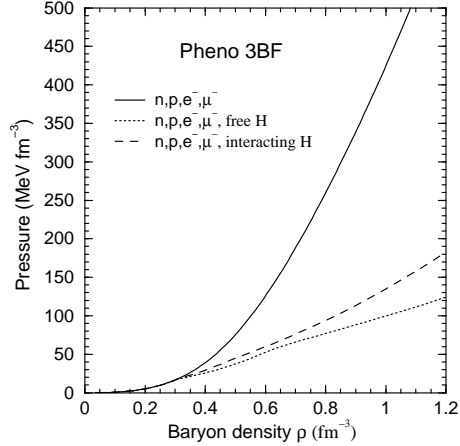}
\caption[]{
The EOS for hyperon-free (solid line) and hyperon-rich (dashed lines) matter.}
\label{cortona2}
\end{figure} 

The resulting EOS is displayed in Fig.~\ref{cortona2}. 
The upper curves show the EOS when stellar matter is composed 
only of nucleons and leptons. 
The inclusion of hyperons (lower curves) produces a much softer EOS,
which turns out to be very similar to the one obtained without TBF.
This is quite astonishing because, in the pure nucleon case, the repulsive
character of TBF at high density increases the stiffness of the EOS,
thus changing dramatically the equation of state. 
However, when hyperons are included, the presence of TBF among nucleons 
enhances the population of $\Sigma^-$ and $\Lambda$ because of the increased 
nucleon chemical potentials with respect to the
case without TBF, thus decreasing the nucleon population.
Of course, this scenario could partly change if hyperon-hyperon 
interactions were known or if TBF would be included also for hyperons, 
but this is beyond our current knowledge of the strong interaction.   

The consequences for the structure of the neutron stars are illustrated  
in Fig.~\ref{f:mrh}, where we display the resulting neutron star 
mass-radius curves, comparing now results obtained with different 
nucleonic TBF, in analogy to Fig.~\ref{f:mr}.
One notes that while in Fig.~\ref{f:mr} the different TBF still yield
quite different maximum masses,
the presence of hyperons equalizes the results, 
leading now to a maximum mass of less than 1.3 solar masses
for all the nuclear TBF.

\begin{figure}[t] 
\includegraphics[scale=0.64,angle=270,bb=130 120 390 120]{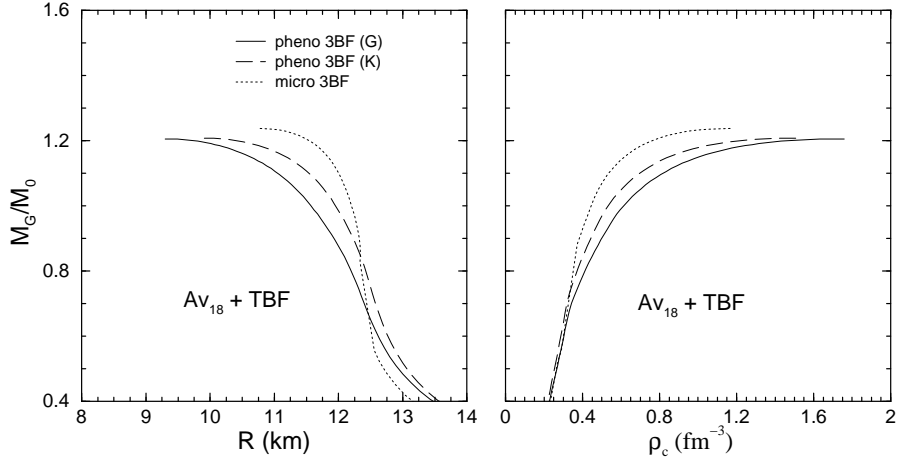}
\caption[]{
Neutron star gravitational mass vs.~radius (left panel) 
and the central baryon density $\rho_c$ (right panel).
Calculations involving different nucleonic TBF are compared.}
\label{f:mrh}
\end{figure} 

This surprising result is due to the strong softening of the baryonic
EOS when including hyperons as additional degrees of freedom,
and we do not expect substantial changes when introducing refinements
of the theoretical framework, 
such as 
hyperon-hyperon potentials, hyperonic TBF, relativistic corrections, etc. 
The only remaining possibility in order to reach larger maximum masses
appears the transition to another phase of dense (quark) matter
inside the star. 
This will be discussed in the following.

\section{Quark matter}

The results obtained with a purely baryonic EOS call for an estimate of
the effects due to the hypothetical presence of quark matter in the interior
of the neutron star.
Unfortunately, the current theoretical description of quark matter 
is burdened with large uncertainties, seriously limiting the 
predictive power of any theoretical approach at high baryonic density. 
For the time being we can therefore only resort
to phenomenological models for the quark matter EOS 
and try to constrain them as well as possible 
by the few experimental information on high density baryonic matter.

One important condition is constituted by the fact 
that certainly in symmetric nuclear matter no phase transition is observed  
below $\approx 3\rho_0$.
In fact some theoretical interpretation of the heavy ion experiments 
performed at the CERN SPS \cite{cern}
points to a possible phase transition at a critical density
$\rho_c \approx 6\rho_0 \approx 1/{\rm fm}^3$.
We will in the following take this value for granted and use an extended 
MIT bag model \cite{chodos}
(requiring a density dependent bag ``constant'')
that is compatible with this condition.

We first review briefly the description of the bulk properties of 
uniform quark matter,
deconfined from the $\beta$-stable hadronic matter mentioned in the
previous section, by using the MIT bag model \cite{chodos}.
The thermodynamic potential of $f=u, d, s$ quarks 
can be expressed as a sum of the kinetic term
and the one-gluon-exchange term \cite{quark,fahri}
proportional to the QCD fine structure constant $\alpha_s$, 
{\bea
 \Omega_f(\mu_f) &=& -{3m_f^4 \over 8\pi^2} \bigg[ 
 {y_f x_f \over 3} \left(2x_f^2-3\right) + \ln(x_f+y_f) \bigg] 
\nonumber \\
 && + \alpha_s{3m_f^4\over 2\pi^3} \bigg\{
 \Big[ y_f x_f - \ln(x_f+y_f) \Big]^2 - {2\over3} x_f^4 + \ln(y_f) 
\nonumber \\
 &&\hskip18mm + 2\ln\Big( {\sigma_{\rm ren} \over m_f y_f} \Big) \Big[ 
 y_f x_f - \ln(x_f + y_f) \Big] \bigg\} \:, \phantom{aaa}
\eea}
where $m_f$ and $\mu_f$ are the $f$ current quark mass and chemical potential, 
respectively, 
$y_f = \mu_f/m_f$, $x_f = \sqrt{y_f^2-1}$,
and
$\sigma_{\rm ren}$ = 313~MeV
is the renormalization point. 
The number density $\rho_f$ of $f$ quarks is related to $\Omega_f$ via
\be
 \rho_f = - {\partial\Omega_f \over \partial\mu_f} \:,
\ee
and the total energy density for the quark system is written as
\be
 \eps_Q(\rho_u,\rho_d,\rho_s) = 
 \sum_f \big( \Omega_f + \mu_f \rho_f \big) + B \:,  
\label{e:eosqm}
\ee
where $B$ is the energy density difference between 
the perturbative vacuum and the true vacuum, {\rm i.e.}, the bag ``constant.''

In the original MIT bag model the bag constant 
$B\approx 55\;\rm MeV\,fm^{-3}$ is used,
while values
$B\approx 210\;\rm MeV\,fm^{-3}$ 
are estimated from lattice calculations \cite{satz1}. 
In this sense $B$ can be considered as a free parameter.
We found, however, that a bag model involving a constant
(density independent) bag parameter $B$,
combined with our BHF hadronic EOS,
will not yield the required phase transition in symmetric matter at
$\rho_c \approx 6\rho_0 \approx 1/{\rm fm}^3$ \cite{qm}.
This can only be accomplished by introducing a density dependence 
of the bag parameter.
(The dependence on asymmetry is neglected at the current level of
investigation).
In practice we use a gaussian parametrization,
\be
 {B(\rho)} 
 = B_\infty + (B_0 - B_\infty)  
 \exp\left[-\beta\Big({\rho \over \rho_0}\Big)^2 \right] 
\ee
with $B_\infty = 50\;\rm MeV\,fm^{-3}$, $B_0 = 400\;\rm MeV\,fm^{-3}$,
and $\beta=0.17$, 
displayed in Fig.~\ref{f:b}(a).

\begin{figure}[t] 
\includegraphics[scale=0.55,angle=270,bb=154 36 385 36]{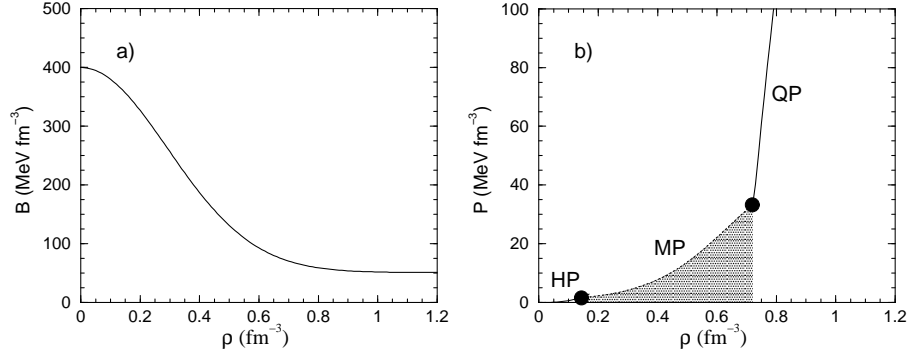}
\caption{
(a) Bag constant $B$ versus baryon number density. 
(b) EOS including both hadronic and quark components.
The shaded region, bordered by two dots, indicates the mixed phase (MP) 
of quarks and hadrons, while HP and QP label the pure hadron and quark phases.}
\label{f:b}
\end{figure} 

For the description of a pure quark phase inside the neutron star,
as for neutrino-free baryonic matter,
the equilibrium equations for the chemical potentials,
\be
 \mu_d = \mu_s = \mu_u + \mu_e \:,
\ee
must be supplemented with the charge neutrality condition 
and the total baryon number conservation,
\bea
 0 &=& {1\over 3}(2\rho_u - \rho_d - \rho_s) - \rho_e \:,
\\
 \rho &=& {1\over3} (\rho_u + \rho_d + \rho_s) \:, 
\eea
in order to determine 
the composition $\rho_f(\rho)$ and the pressure of the quark phase,
\be
 P_Q(\rho) = \rho {d\eps_Q \over \ d\rho} - \eps_Q \:.
\ee

However, a more realistic model for  
the phase transition between baryonic and quark phase inside the star
is the Glendenning construction \cite{glen},
which determines the range of baryon density where both phases coexist. 
The essential point of
this procedure is that both the hadron and the quark phase are allowed to be 
separately charged, still preserving the total charge neutrality. 
This implies that neutron star matter can be treated as a two-component
system, and therefore can be parametrized by two chemical potentials
like electron and baryon chemical potentials $\mu_e$ and $\mu_n$. 
The pressure is the same in the two phases to ensure mechanical stability, 
while the chemical potentials of
the different species are related to each other satisfying chemical and
beta stability. 
The Gibbs condition for mechanical and chemical equilibrium 
at zero temperature between both phases reads
\be
 p_H(\mu_e,\mu_n) = p_Q(\mu_e,\mu_n) = p_M(\mu_n) \:. 
\label{e:mp}
\ee
From the intersection of the two surfaces representing the hadron and 
the quark phase one can calculate the equilibrium chemical potentials 
of the mixed phase, as well as 
the charge densities $\rho_c^H$ and $\rho_c^Q$ 
and therefore the volume fraction $\chi$ occupied 
by quark matter in the mixed phase,
\be
 \chi = { \rho_c^H \over \rho_c^H - \rho_c^Q } \:.
\label{e:chi}
\ee
From this, the baryon density $\rho_M$
and the energy density $\eps_M$
of the mixed phase can be calculated as
\bea
 \rho_M &=& \chi \rho_Q + (1 - \chi) \rho_H \:, 
\\
 \eps_M &=& \chi \eps_Q + (1 - \chi) \eps_H \:. 
\label{e:mp1}
\eea
The EOS resulting from this procedure is shown in Fig.~\ref{f:b}(b),
where the pure hadron, mixed, and pure quark matter portions are indicated.
The mixed phase begins actually at a quite low density around $\rho_0$.
Clearly the outcome of the mixed phase construction might be substantially changed, 
if surface and Coulomb energies were taken into account \cite{surf}. 
For the time being these are, however, unknown and have been neglected.

\begin{figure}[t] 
\includegraphics[scale=0.63,angle=270,bb=150 50 390 50]{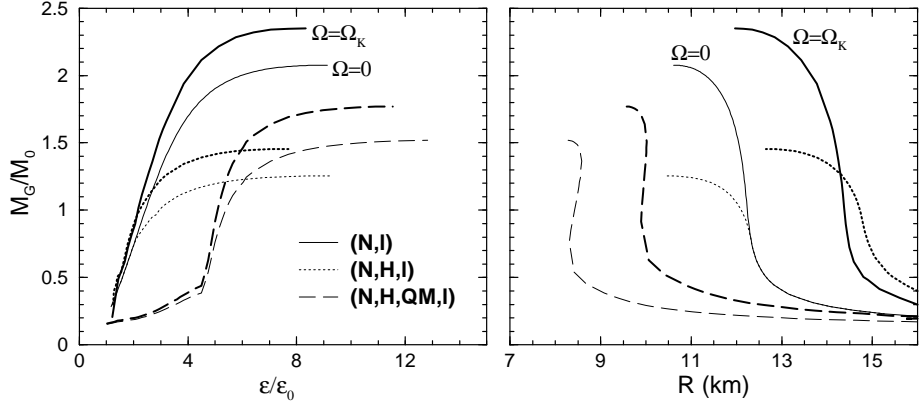}
\caption{
The gravitational mass (in units of the solar mass $M_\odot$) 
versus the normalized central energy density 
($\epsilon_0=156\;\rm MeV\,fm^{-3}$) (left panel) 
and versus the equatorial radius (right panel). 
The thin lines represent static equilibrium configurations, 
whereas the thick lines display configurations
rotating at their respective Kepler frequencies. 
Several different stellar matter compositions are considered 
(see text for details).}
\label{f:mrq}
\end{figure} 

The final result for the structure of hybrid neutron stars 
is shown in Fig.~\ref{f:mrq},
displaying mass-radius and mass-central density relations.
It is evident that the most striking effect of the inclusion of quark
matter is the increase of the maximum mass, now reaching about 
$1.5\;M_\odot$.
At the same time, the typical neutron star radius is reduced by about
3 km to typically 9 km. 
Hybrid neutron stars are thus more compact than purely hadronic ones
and their central energy density is larger.
For completeness, the figure shows 
besides static neutron star configurations
also those rotating at the maximum (Kepler) frequency \cite{qmrot}.
In that case one observes a further enhancement of the maximum mass
to about $1.8\;M_\odot$,
and an increase of the typical equatorial radius by about 1 km.

\begin{figure}[t] 
\includegraphics[scale=0.6,angle=270,bb=150 100 500 100]{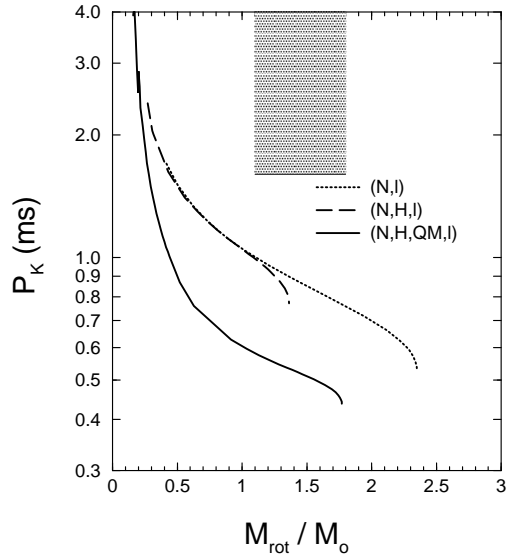}
\caption{
Kepler period versus the rotational mass for
purely hadronic stars as well as hybrid stars.  
The following core compositions are considered: 
i) nucleons and leptons (dotted line); 
ii) nucleons, hyperons, and leptons (dashed line); 
iii) hadrons, quarks, and leptons (solid line). 
The shaded area represents the current range of observed data.}
\label{f:kp}
\end{figure} 

Finally, in Fig.~\ref{f:kp} we display the Kepler periods 
$P_K$ ($= 2\pi /\Omega_K$)
versus the rotational star mass for several different stellar
sequences based on different EOS. 
Purely hadronic stars,
shown by the dotted and long-dashed lines respectively, 
show instability against mass shedding first, 
because of their relatively large equatorial radii.  
Their limiting mass configurations are characterized by values of the 
Kepler period larger than half a millisecond, 
in agreement with results usually found in the literature \cite{pulsar}. 
In contrast, hybrid stars can reach stable periods
smaller than half a millisecond.

\section{Conclusions}

In this contribution we reported the theoretical description of nuclear
matter in the BHF approach and its various refinements, 
with the application to neutron star structure calculation.
We pointed out the important role of TBF at high density,
which is, however, strongly compensated by the inclusion of hyperons.
The resulting hadronic neutron star configurations have maximum masses 
of only about $1.3\,M_\odot$,
and the presence of quark matter inside the star 
is required in order to reach larger values.

Concerning the quark matter EOS,
we found that a density dependent bag parameter $B(\rho)$ is
necessary in order to be compatible with  
the CERN-SPS findings on the phase transition
from hadronic to quark matter. 
Joining the corresponding EOS with the baryonic one, 
maximum masses of about $1.6\;M_\odot$ are reached,
in line with other recent calculations of neutron star properties employing
various phenomenological RMF nuclear EOS together with either effective mass bag
model \cite{bag} or Nambu-Jona-Lasinio model \cite{njl} EOS for quark matter.

The value of the maximum mass of neutron stars obtained according to
our analysis appears rather robust with respect to the uncertainties
of the nuclear and the quark matter EOS.
Therefore, the experimental observation of a very heavy
($M \gtrsim 1.6 M_\odot$) neutron star, 
as claimed recently by some groups \cite{kaaret} 
($M \approx 2.2\;M_\odot$), 
if confirmed, would suggest that 
either serious problems are present for the current theoretical modelling
of the high-density phase of nuclear matter,
or that the assumptions about
the phase transition between hadron and quark phase
are substantially wrong. 
In both cases, one can expect a well defined hint on the
high density nuclear matter EOS.

We would like to thank our collaborators 
J. Cugnon, A. Lejeune, U. Lombardo, F. Mathiot, 
P.K. Sahu, F. Weber, X.R. Zhou, and W. Zuo.



\begin{chapthebibliography}{1}
\addcontentsline{toc}{section}{References}

\bibitem{bon} 
 P. Bonche, E. Chabanat, P. Haensel, J. Meyer, and R. Schaeffer, 
 Nucl. Phys. {\bf A643}, 441 (1998).

\bibitem{wal} 
 B. D. Serot and J. D. Walecka, 
 Adv. Nucl. Phys. {\bf 16}, 1 (1986).

\bibitem{bal} 
 M. Baldo, 
 {\it The many body theory of the nuclear equation of state}
 in Nuclear Methods and the Nuclear Equation of State, 1999,
 Ed. M. Baldo, World Scientific, Singapore.

\bibitem{dbhf}
 R. Machleidt, 
 Adv. Nucl. Phys. {\bf 19}, 189 (1989) and references quoted therein;
 G. Q. Li, R. Machleidt, and R. Brockmann,
 Phys. Rev. {\bf C45}, 2782 (1992).

\bibitem{pan} 
 A. Akmal and V. R. Pandharipande, 
 Phys. Rev. {\bf C56}, 2261 (1997);
 A. Akmal, V. R. Pandharipande, and D. G. Ravenhall,
 Phys. Rev. {\bf C58}, 1804 (1998);
 J. Morales, V. R. Pandharipande, and D. G. Ravenhall,
 Phys. Rev. {\bf C66}, 054308 (2002).

\bibitem{chi} 
 N. Kaiser, S. Fritsch, and W. Weise, 
 Nucl. Phys. {\bf A697}, 255 (2002).

\bibitem{gle} 
 N. K. Glendenning,
 Nucl. Phys. {\bf A493}, 521 (1989);
 {\em Compact Stars, Nuclear Physics, Particle Physics, and General Relativity},
 2nd ed., 2000, Springer-Verlag, New York.

\bibitem{son} 
 H. Q. Song, M. Baldo, G. Giansiracusa, and U. Lombardo,
 Phys. Rev. Lett. {\bf 81}, 1584 (1998).

\bibitem{thl}
 M. Baldo, A. Fiasconaro, H. Q. Song, G. Giansiracusa, and U. Lombardo,
 Phys. Rev. {\bf C65}, 017303 (2002).

\bibitem{gra} 
 P. Grang\a'e, A. Lejeune, M. Martzolff, and J.-F. Mathiot,
 Phys. Rev. {\bf C40}, 1040 (1989).

\bibitem{mic}
 S. A. Coon, M. D. Scadron, P. C. McNamee, B. R. Barrett, D. W. E. Blatt,
 and B. H. J. McKellar,
 Nucl. Phys. {\bf A317}, 242 (1979).

\bibitem{lom} 
 A. Lejeune, U. Lombardo, and W. Zuo,
 Phys. Lett. {\bf B477}, 45 (2000);
 W. Zuo, A. Lejeune, U. Lombardo, and J.-F. Mathiot,
 Nucl. Phys. {\bf A706}, 418 (2002);
 Eur. Phys. Journ. {\bf A14}, 469 (2002).

\bibitem{hypmat} 
 H.-J. Schulze, A. Lejeune, J. Cugnon, M. Baldo, and U. Lombardo,
 Phys. Lett. {\bf B355}, 21 (1995);
 Phys. Rev. {\bf C57}, 704 (1998).

\bibitem{hypnuc} 
 J. Cugnon, A. Lejeune, and H.-J. Schulze,
 Phys. Rev. {\bf C62}, 064308 (2000);
 I. Vida\~na, A. Polls, A. Ramos, and H.-J. Schulze,
 Phys. Rev. {\bf C64}, 044301 (2001).

\bibitem{hypns} 
 M. Baldo, G. F. Burgio, and H.-J. Schulze,
 Phys. Rev. {\bf C58}, 3688 (1998);
 Phys. Rev. {\bf C61}, 055801 (2000).

\bibitem{barc} 
 I. Vida\a~na, A. Polls, A. Ramos, L. Engvik, and M. Hjorth-Jensen, 
 Phys. Rev. {\bf C62}, 035801 (2000).

\bibitem{brow} 
 G. E. Brown et al., 
 Comm. Nucl. Part. Phys. {\bf 17}, 39 (1987).

\bibitem{land}
 W. Zuo, Caiwan Shen, and U. Lombardo, 
 Phys. Rev. {\bf C67}, 037301 (2003).

\bibitem{uix}
 B. S. Pudliner, V. R. Pandharipande, J. Carlson, and R. B. Wiringa,
 Phys. Rev. Lett. {\bf 74}, 4396 (1995).

\bibitem{v18}
 R. B. Wiringa, V. G. J. Stoks, and R. Schiavilla,
 Phys. Rev. {\bf C51}, 38 (1995).

\bibitem{ill}
 S. C. Pieper, V. R. Pandharipande, R. B. Wiringa, and J. Carlson,
 Phys. Rev. {\bf C64}, 014001 (2001).

\bibitem{bom} 
 A. Lejeune, P. Grang\a'{e}, M. Martzolff, and J. Cugnon,
 Nucl. Phys. {\bf A453}, 189 (1986);
 I. Bombaci and U. Lombardo, 
 Phys. Rev. {\bf C44}, 1892 (1991);
 W. Zuo, I. Bombaci, and U. Lombardo, 
 Phys. Rev. {\bf C60}, 024605 (1999).

\bibitem{ns}
 S. Shapiro and S. A. Teukolsky,
 {\em Black Holes, White Dwarfs, and Neutron Stars},
 1983, ed. John Wiley \& Sons, New York.

\bibitem{aa}
 M. Baldo, I. Bombaci, and G. F. Burgio, 
 Astron. Astroph. {\bf 328}, 274 (1997).

\bibitem{fey}
 R. Feynman, F. Metropolis, and E. Teller,
 Phys. Rev. {\bf C75}, 1561 (1949).

\bibitem{baym}
 G. Baym, C. Pethick, and D. Sutherland,
 Astrophys. J. {\bf 170}, 299 (1971).

\bibitem{nv}
 J. W. Negele and D. Vautherin,
 Nucl. Phys. {\bf A207}, 298 (1973).

\bibitem{qm}
 G. F. Burgio, M. Baldo, P. K. Sahu, A. B. Santra, and H.-J. Schulze,
 Phys. Lett. {\bf B526}, 19 (2002);
 G. F. Burgio, M. Baldo, P. K. Sahu, and H.-J. Schulze,
 Phys. Rev. {\bf C66}, 025802 (2002);
 M. Baldo, M. Buballa, G. F. Burgio, F. Neumann, M. Oertel, and H.-J. Schulze,
 Phys. Lett. {\bf B562}, 153 (2003).

\bibitem{mae89}
 P. Maessen, Th. Rijken, and J. de Swart,
 Phys. Rev. {\bf C40}, 2226 (1989).

\bibitem{cern}
 U. Heinz and M. Jacobs, nucl-th/0002042;
 U. Heinz, Nucl. Phys. {\bf A685}, 414 (2001).

\bibitem{chodos}
 A. Chodos,  R. L. Jaffe, K. Johnson, C. B. Thorn, and V. F. Weisskopf,
 Phys. Rev. {\bf D9}, 3471 (1974).

\bibitem{quark}
 E. Witten, 
 Phys. Rev. {\bf D30}, 272 (1984); 
 G. Baym, E. W. Kolb, L. McLerran, T. P. Walker, and R. L. Jaffe, 
 Phys. Lett. {\bf B160}, 181 (1985); 
 N. K. Glendenning, 
 Mod. Phys. Lett. {\bf A5}, 2197 (1990).

\bibitem{fahri}
 E. Fahri and R. L. Jaffe, 
 Phys. Rev. {\bf D30}, 2379 (1984).

\bibitem{satz1}
 H. Satz,
 Phys. Rep. {\bf 89}, 349 (1982).  

\bibitem{glen}
 N. K. Glendenning,
 Phys. Rev. {\bf D46}, 1274 (1992).

\bibitem{surf}
 T. Tatsumi, M. Yasuhira, and D. N. Voskresensky,
 Nucl. Phys. {\bf A718}, 359 (2003);  {\bf A723}, 291 (2003);
 D. N. Voskresensky, these proceedings.

\bibitem{qmrot}
 G. F. Burgio, H.-J. Schulze, and F. Weber,
 Astron. Astrophys. {\bf 408}, 675 (2003).

\bibitem{pulsar}
 F. Weber, 
 {\em Pulsars as Astrophysical Laboratories for Nuclear and Particle Physics}, 
 Institute of Physics Publishing, Bristol and Philadelphia (1999);
 F. Weber, 
 J. Phys. G: Nucl. Part. Phys. {\bf 25}, R195 (1999).

\bibitem{bag}
 K. Schertler, C. Greiner, P. K. Sahu and M. H. Thoma, 
 Nucl. Phys. {\bf A637}, 451 (1998);
 K. Schertler, C. Greiner, J. Schaffner-Bielich, and M. H. Thoma,
 Nucl. Phys. {\bf A677}, 463 (2000).

\bibitem{njl}
 K. Schertler, S. Leupold, and J. Schaffner-Bielich, 
 Phys. Rev. {\bf C60}, 025801 (1999).

\bibitem{kaaret}
 P. Kaaret, E. Ford, and K. Chen, 
 Astrophys. J. Lett. {\bf 480}, L27 (1997); 
 W. Zhang, A. P. Smale, T. E. Strohmayer, and J. H. Swank,
 Astrophys. J. Lett. {\bf 500}, L171 (1998).

\end{chapthebibliography}
\end{document}